\documentclass[12pt]{article}
\usepackage{graphicx}
\usepackage{url}

\def\t{{\bar t}}

\def\PTavt{p_{\mathrm{T,avt}}}
\def\PTt{p_{\mathrm{T,t}}}
\def\PTtbar{p_{\mathrm{T,{\bar t}}}}
\def\Yavt{y_{\mathrm{avt}}}

\def\as{\alpha_S}

\newcommand\pubnumber{Cavendish-HEP-17/15, \\ CP3-17-55, \\ NIKHEF/2017-065,\\ TUM-HEP-1119/17,\\ TTK-17-46}
\newcommand\pubdate{\today}

\def\institute{ $^{1}$Institut f\"ur Theoretische Teilchenphysik und Kosmologie, \\ RWTH Aachen University, D-52056 Aachen, Germany\\
$^{2}$Cavendish Laboratory, University of Cambridge, Cambridge CB3 0HE, UK\\
$^{3}$Technische Universit\"{a}t M\"{u}nchen, James-Franck-Str.~1, D-85748 Garching, Germany\\
$^{4}$Centre for Cosmology, Particle Physics and Phenomenology (CP3), \\ Universit\'e Catholique de Louvain, Chemin du Cyclotron 2, B-1348 Louvain la Neuve, Belgium\\
$^{5}$Nikhef, Science Park 105, NL-1098 XG Amsterdam, The Netherlands\\
$^{6}$Sorbonne Universit\'es, UPMC Univ. Paris 06,\\ UMR 7589, LPTHE, F-75005, Paris, France\\
$^{7}$CNRS, UMR 7589, LPTHE, F-75005, Paris, France}

\def\Title#1{\begin{center} {\Large #1 } \end{center}}
\def\Author#1{\begin{center}{ \sc #1} \end{center}}
\def\Address#1{\begin{center}{ \it #1} \end{center}}

\newcommand\pubblock{\rightline{\begin{tabular}{l} \pubnumber\\
         \pubdate  \end{tabular}}}
\newenvironment{Abstract}{\begin{quotation}  }{\end{quotation}}
\newenvironment{Presented}{\begin{quotation} \begin{center} 
             PRESENTED AT\end{center}\bigskip 
      \begin{center}\begin{large}}{\end{large}\end{center} \end{quotation}}
\def\Acknowledgements{\bigskip  \bigskip \begin{center} \begin{large}
             \bf ACKNOWLEDGEMENTS \end{large}\end{center}}

\def\beq{\begin{equation}}
\def\eeq#1{\label{#1}\end{equation}}
\def\eeqn{\end{equation}}

\def\beqa{\begin{eqnarray}}
\def\eeqa#1{\label{#1}\end{eqnarray}}
\def\eeqan{\end{eqnarray}}

\let\bar=\overbar

\def\Dslash{\not{\hbox{\kern-4pt $D$}}}
\def\dslash{\not{\hbox{\kern-2pt $\del$}}}

\def\msb{{\bar{\ssstyle M \kern -1pt S}}}

\begin{document}
\begin{titlepage}
\pubblock

\vfill
\Title{Top-quark pair production at NNLO QCD + NLO EW accuracy: Tevatron results }
\vfill
\Author{ Micha\l{}  Czakon$^{1}$, David Heymes$^{2}$, Alexander Mitov$^{2}$, Davide Pagani$^{3}$, Ioannis Tsinikos$^{3,4}$, Marco Zaro$^{5,6,7}$}
\Address{\institute}
\vfill
\begin{Abstract}
We calculate all main top-quark pair differential distributions at the Tevatron. For the first time Tevatron predictions for this process include all Standard Model corrections through NLO (referred to as {\it complete-NLO}) consistently combined with previously computed NNLO QCD corrections. We also assess, for the first time, the impact of dynamical scales on predictions for the latest Tevatron measurements.
\end{Abstract}
\vfill
\begin{Presented}
$10^{th}$ International Workshop on Top Quark Physics\\
Braga, Portugal,  September 17--22, 2017\\
by Davide Pagani
\end{Presented}
\vfill
\end{titlepage}
\def\thefootnote{\fnsymbol{footnote}}
\setcounter{footnote}{0}

\section{Introduction}

In these proceedings we provide precise predictions for top-quark pair differential distributions at the Tevatron proton--antiproton collider with c.m.~energy of 1.96 TeV. For the first time, NNLO QCD results with dynamical scales are calculated for the Tevatron and electroweak (EW) corrections are also taken into account.  The results presented here are new and have not been included in other publications. The calculation framework is identical to the one described in detail in ref.~\cite{Czakon:2017wor}; in the following we recall its salient features.

Our predictions fully account for NNLO QCD corrections at $\mathcal O(\alpha_s^4)$ as well as for the {\it complete-NLO} ones, {\it i.e.}, the complete set of Standard Model corrections at NLO. For $t \bar t$ production the latter corresponds to the NLO QCD,  the NLO EW  as well as the subleading NLO contributions at $\mathcal O(\alpha^2 \as)$ and $\mathcal O(\alpha^3)$ together with the LO ones at $\mathcal O(\alpha \as)$ and  $\mathcal O(\alpha^2)$. The QED and exact purely weak components of the NLO EW corrections were separately calculated for the first time in Refs.~\cite{Hollik:2007sw} and \cite{Kuhn:2006vh,Bernreuther:2006vg}, respectively. 
The NNLO QCD corrections have been computed following the approach of refs.~\cite{Czakon:2015owf,Czakon:2016dgf}, extending the earlier fixed-scale NNLO QCD Tevatron calculation~\cite{Czakon:2016ckf} by including dynamical scales. The complete-NLO corrections have been calculated with a private extension of the {\sc\small MadGraph5\_aMC@NLO} code \cite{Alwall:2014hca,Frixione:2015zaa,Pagani:2016caq} that has already been successfully employed also in refs.~\cite{Frederix:2016ost, Frederix:2017wme}. As discussed and motivated in ref.~\cite{Czakon:2017wor}, NNLO QCD and NLO EW corrections are combined in the multiplicative approach. For this reason we denote our predictions as ``QCD$\times$EW''.

Besides the different center-of-mass energy  and the fact that proton--antiproton collisions are considered, the input parameters are the same as the ones in ref.~\cite{Czakon:2017wor}. In particular, we use the {\sc\small LUXqed\_\-plus\_PDF4LHC15\_nnlo\_100} PDF set, which is based on the {\sc \small PDF4LHC} set \cite{Butterworth:2015oua, Ball:2014uwa, Harland-Lang:2014zoa, Dulat:2015mca} and exploits the strategy of refs.~\cite{Manohar:2016nzj,Manohar:2017eqh} for the determination of the photon distribution.  Moreover, we provide predictions obtained with both dynamical and fixed scales. While the use of the former should be preferred, results obtained with the latter can be easily compared with existing predictions. For the static, fixed-scale choice we use the top-quark mass ($\mu=m_t$) while for the dynamical scale we employ the functional forms discussed in ref.~\cite{Czakon:2016dgf}:
\begin{eqnarray}
\mu &=& \frac{m_{T,t}}{2}~~{\rm for~the} ~ \PTt ~ {\rm distribution}, \label{eq:scalemT}\\
\mu &=& \frac{m_{T,\t}}{2}~~{\rm for~the} ~ \PTtbar ~ {\rm distribution}, \label{eq:scalemTbar}\\
\mu &=& \frac{H_T}{4} = \frac{1}{4} \left( m_{T,t} + m_{T,\bar t} \right)~~{\rm for~all~other~distributions}.
\label{eq:scaleHT}
\end{eqnarray}

Theoretical uncertainties due to missing higher orders are estimated via the 7-point variation of $\mu_r$ and $\mu_f$ in the interval $\{\mu/2,2\mu\}$ with $1/2\leq\mu_r/\mu_f\leq2$.

\section{Phenomenological predictions for the Tevatron at 1.96 TeV}
\begin{figure}[t]
\centering
\includegraphics[width=0.48\textwidth]{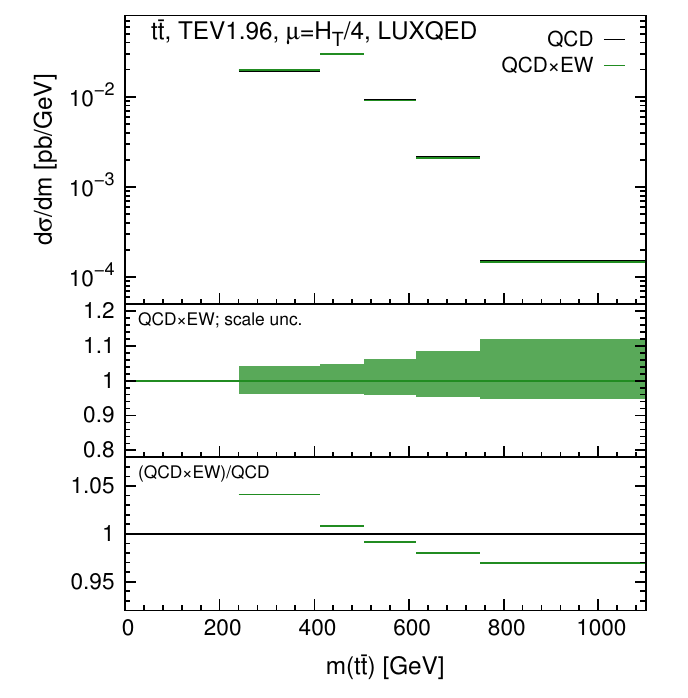}
\includegraphics[width=0.48\textwidth]{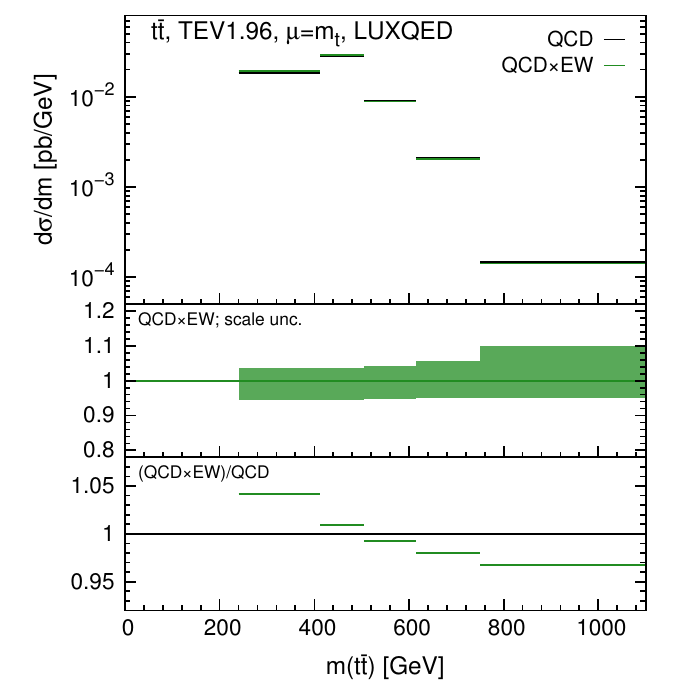}
\caption{Differential predictions for the top-pair invariant mass $m(t\t)$. The left (right) plot shows results obtained with dynamical (static) scales.}
\label{fig:mtt}
\end{figure}
\begin{figure}[t]
\centering
\includegraphics[width=0.48\textwidth]{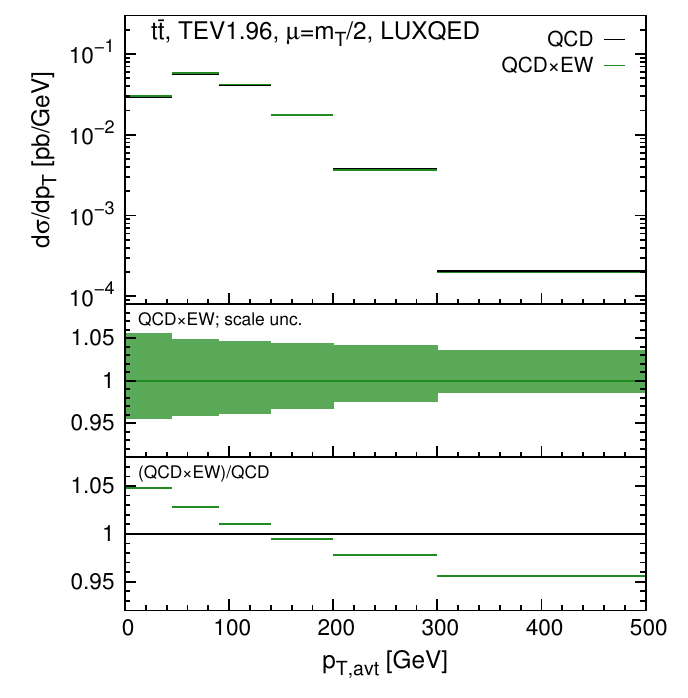}
\includegraphics[width=0.48\textwidth]{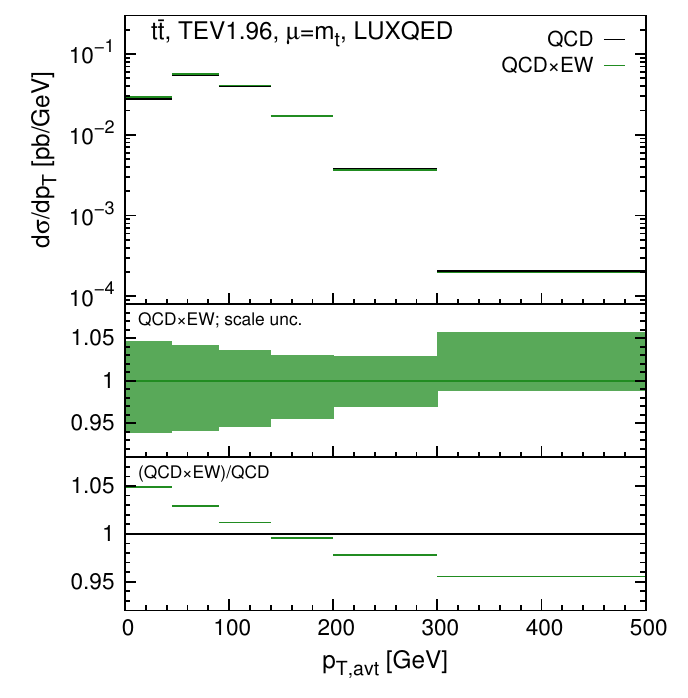}
\caption{ Same as Fig.~\ref{fig:mtt} but for the top-pair average transverse momentum $\PTavt$. 
}
\label{fig:PTavt}
\end{figure}
\begin{figure}[t]
\centering
\includegraphics[width=0.48\textwidth]{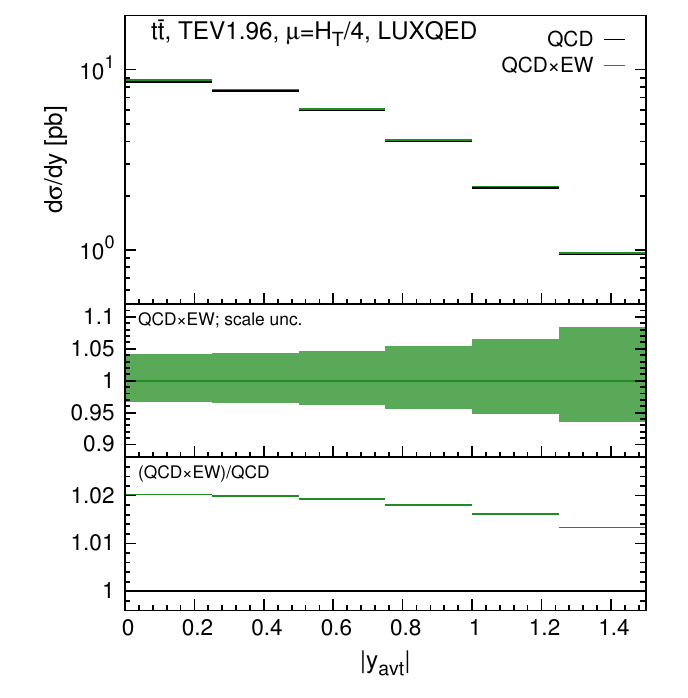}
\includegraphics[width=0.48\textwidth]{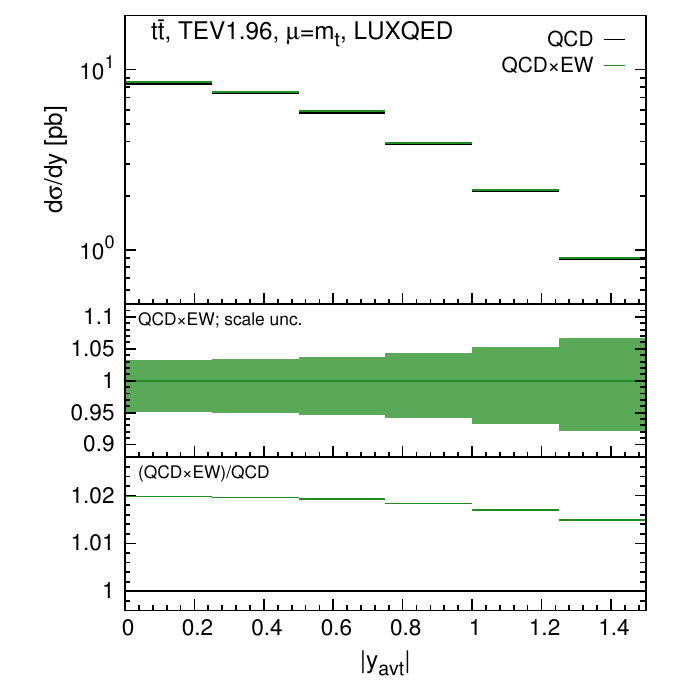}
\caption{Same as Fig.~\ref{fig:mtt} but for for the top-pair average absolute rapidity $|\Yavt|$.
}
\label{fig:Yavt}
\end{figure}
We provide phenomenological predictions for the following distributions: the top-pair invariant mass $m(t\t)$, the average of the transverse momentum ($\PTavt$) and absolute rapidity ($|\Yavt|$)  of the top quark and antiquark which we calculate {\it not} on an event-by-event basis but by averaging the corresponding distributions. The $m(t\t)$, $\PTavt$ and $|\Yavt|$ distributions are shown, respectively, in figures \ref{fig:mtt}-\ref{fig:Yavt}; 
the plots on the left (right) are computed with dynamical (fixed) scales. Each plot has the following layout: in the main frame we show the absolute predictions for the considered observable  both at ``QCD$\times$EW'' accuracy (green curve) and NNLO-QCD accuracy (black curve). We dub the pure NNLO QCD calculation ``QCD'' and recall that it has already been studied in ref.~\cite{Czakon:2016ckf}.  The first inset contains the scale-uncertainty band of the QCD$\times$EW prediction, while in the second inset we show the bin-by-bin ratio of the green and black curves of the main frame, {\it i.e.},  the relative effect of EW corrections w.r.t.~the NNLO QCD predictions.
 
From Figure~\ref{fig:mtt} it can be appreciated that except for the last bin the scale uncertainty of the $m(t \bar t)$ distribution is well below ±10\%. The EW corrections are comparable in size to the QCD scale uncertainty. They lead to a non-trivial shape modification, being
positive (+5\%) at low values of the invariant mass and negative (-3\%) for large ones (due to the Sudakov suppression).

As can be seen in Figure~\ref{fig:PTavt},  at variance with the case of the invariant mass distribution, scale uncertainties decrease for larger values of $\PTavt$, and never exceed 6\%. The effect of EW corrections is instead similar to the case of the invariant mass. Only in the last $\PTavt$ bin they exceed in absolute value the scale uncertainty band. 

Finally, in Figure~\ref{fig:Yavt} we show the $|\Yavt|$ distribution. Since this observable is dominated by events at small-to-medium values of the transverse momentum, the effect of EW corrections is rather small (1-2\%), while scale uncertainties are between 3 and 7\%, growing with rapidity. If we compare the scale uncertainty bands and the effect of EW corrections between predictions with fixed and dynamical scales, we generally observe very little differences. We remind the reader that similarly to the case of the transverse momentum, the {\it absolute} rapidity distributions of top and antitop are equal; on the other hand the rapidity distributions of the top and antitop quark differ in $p\bar p$ collisions and it is this difference that leads to the so-called forward-backward $t\bar t$ asymmetry. Both QCD \cite{Kuhn:1998jr,Czakon:2014xsa} and EW \cite{Hollik:2011ps} radiative effects are very large for this observable; updated results have recently been presented in ref.~\cite{Czakon:2017lgo}.

\begin{figure}[t]
\centering
\includegraphics[width=0.48\textwidth]{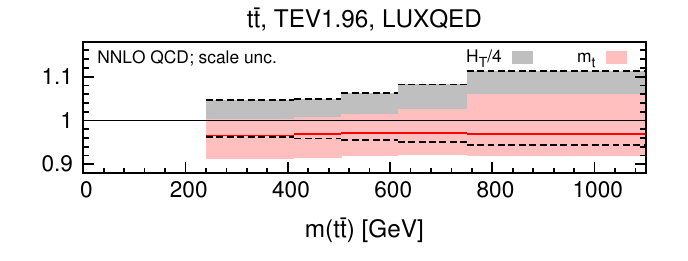}
\includegraphics[width=0.48\textwidth]{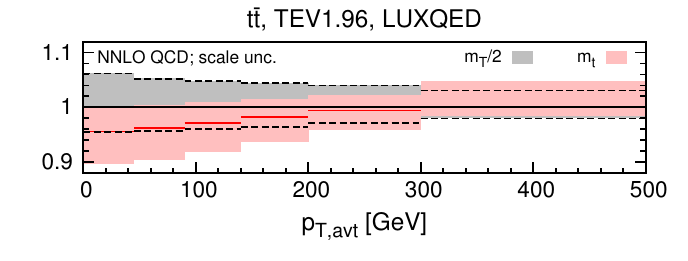}\\
\includegraphics[width=0.48\textwidth]{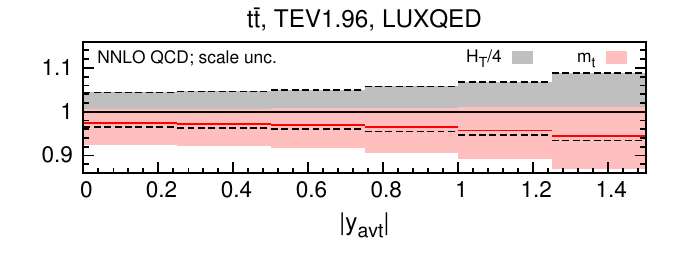}
\caption{Comparison between the dynamical (black) and static (red) scale choice for $m(t\t)$, $\PTavt$ and $|\Yavt|$.}
\label{fig:ratio}
\end{figure}

It is interesting to note that while dynamical scales slightly reduce the scale-un\-cer\-tain\-ty band for $\PTavt$ and $|\Yavt|$, they increase it for $m(t\t)$.  
A better insight of the comparison between static and dynamical scales can be obtained by looking at Figure~\ref{fig:ratio}, where the two choices (fixed scale in red, dynamical scale in black) are compared for the three observables considered. Both fixed-scale and dynamical-scale bands are normalised to the central value of the predictions based on dynamical scales. It can be appreciated that the difference between the two scale choices is always below $\sim$5\%, demonstrating the convergence of the perturbative series and supporting the reliability of the scale-variation based uncertainty estimates.

\section{Conclusion}
In these proceedings we have presented differential distributions for top-pair production at the Tevatron, obtained by combining the NNLO QCD corrections with the complete-NLO ones. These results are shown here for the first time and have not been included in other publications. We have provided results with both fixed and dynamical scales. Our results show that the impact of EW corrections is comparable to, or larger than, the residual uncertainty due to scale variation which is, however, quite small at NNLO. 

All numerical results presented in these proceedings are available in electronic format at the web repository:
\\
\\
\url{http://www.precision.hep.phy.cam.ac.uk/results/ttbar-nnloqcd-nloew}

\Acknowledgements
A.M. thanks the Department of Physics at Princeton University for hospitality during the completion of this work. The work of M.C. is supported in part by grants of the DFG and BMBF. The work of D.H. and A.M. is supported by the UK STFC grants ST/L002760/1 and ST/K004883/1 and by the European Research Council Consolidator Grant ``NNLOforLHC2". The work of D.P is supported by the Alexander von Humboldt Foundation, in the framework of the Sofja Kovalevskaja Award Project ``Event Simulation for the Large Hadron Collider at High Precision''. The work of I.T. is supported by the F.R.S.-FNRS ``Fonds de la Recherche Scientifique'' (Belgium). The work of M.Z. has been supported by the Netherlands
National Organisation for Scientific Research (NWO), by the European Union's Horizon 2020 research and
innovation programme under the Marie Sklodovska-Curie grant
agreement No 660171 and in part by the ILP LABEX (ANR-10-LABX-63),
in turn supported by French state funds managed by the ANR
within the ``Investissements d'Avenir'' programme
under reference ANR-11-IDEX-0004-02.

\end{document}